\documentstyle[aps,epsfig,preprint]{revtex}

\newcommand{\be}{\begin{equation}}
\newcommand{\ee}{\end{equation}}
\newcommand{\bea}{\begin{eqnarray}}
\newcommand{\eea}{\end{eqnarray}}
\newcommand{\bsa}{\begin{subeqnarray}}
\newcommand{\esa}{\end{subeqnarray}}

\newcommand{\Veff}%
{{\cal V}^{\mbox{\scriptsize p}}_{\mbox{\scriptsize eff}}}

\begin{document}
\tighten

\title{Microscopic three-body force for asymmetric nuclear matter}

\date{Preliminary draft: \today}

\author{W. Zuo}
 \address{Institute of Modern Physics, Chinese Academy of Sciences,
         730000 Lanzhou, China}

\author{A. Lejeune}
 \address{Institut de Physique, B5 Sart-Tilman, B-4000 Li\`ege 1, Belgium}

\author{U. Lombardo$^1$}
 \address{Dipartimento di Fisica, 57 Corso Italia, and
          INFN-LNS, 44 Via Santa Sofia 95125 Catania, Italy }
\author{J. F. Mathiot}
 \address{Laboratoire de Physique Corpusculaire, Universit\'e Blaise-Pascal,
           CNRS-IN2P3, 24 Avenue des Landais, F-63177 Aubiere Cedex, France }

\maketitle
\begin{abstract}

Brueckner calculations including a microscopic three-body force have 
been extended to isospin asymmetric nuclear matter. The effects of 
the three-body force on the equation of state and on the 
single-particle properties of nuclear matter are discussed with a view to
possible applications in nuclear physics and astrophysics. 
It is shown that, even in the presence of the three-body force,
the empirical parabolic law of the energy per nucleon vs isospin 
asymmetry $\beta=(N-Z)/A$ is fulfilled in the whole asymmetry range 
$0\le\beta\le 1$  up to high densities. 
The three-body force provides a strong enhancement of symmetry energy 
increasing with the density in good agreement with relativistic approaches. 
The Lane's assumption that proton and neutron mean fields linearly vary vs 
the isospin parameter is violated at high 
density in the presence of the three-body force. Instead the momentum 
dependence of the mean fields is rather insensitive to three body force which 
brings about a linear isospin deviation of the neutron and proton effective
masses.  The isospin effects on multifragmentation events and collective flows 
in heavy-ion collisions are briefly discussed along with the conditions for 
direct URCA processes to occur in the neutron-star cooling. 

\centerline{PACS numbers: 25.70-z, 13.75.Cs, 21.65.+f, 24.10.Cn}

\centerline{Keywords: three body force, asymmetric nuclear matter, isospin}

\centerline{$^1$ e-mail: lombardo@lns.infn.it}
\end{abstract}

\newpage
\normalsize

 {\bf 1.~Introduction}

\par
The equation of state (EOS) of neutron-rich matter is a source of important 
theoretical predictions on the properties of neutron stars, heavy-ion 
collisions (HIC) and nuclei at the neutron drip line \cite{BALD}. The 
interest has been mainly focussed on the properties of the symmetry energy, 
including its dependence 
upon the baryonic density. Studies of neutron-star cooling \cite{LATT}, 
spin-polarized states \cite{MARG}, collective flows and 
isospin distillation in HIC \cite{BAO2,XU,DITO3} have 
stressed that those phenomena are very sensitive to the values of the   
symmetry energy in the respective density domains.   
The EOS of isospin-asymmetric nuclear matter has been recently studied in 
the framework of the Brueckner-Bethe-Goldstone (BBG) theory \cite{BOMB,ASI}. 
The convergence of the BBG hole-line expansion has been  
assessed in recent times with high accuracy \cite{BALDO}, even if the 
the saturation properties of nuclear matter are not reproduced. It is commonly 
recognized that the missing saturation is due to the model of nucleons as 
structureless particles interacting via a bare two-body force (2BF) and one 
has to introduce three-body
forces (3BF). The first microscopic model of 3BF was the Fujita-Miyazawa 
model \cite{FUJI} where the isobar $\Delta(1232)$ is excited in a 
pion exchange interaction between two nucleons. 
The model was later extended to the $N(1440)$ Roper resonance \cite{MATH}. 
It has been also recognized that the main relativistic effect introduced by 
Dirac-Brueckner approach is the excitation of the negative energy states of the
Fermi sea which can be described in terms of a 3TB \cite{BROWN} as well.  

The above mentioned 3BF components have a strong saturating effect as
already shown by a Brueckner-Hartree-Fock (BHF) calculation with the Paris 
potential as 2BF \cite{MATH}. This 3BF has been re-adopted in Ref.~\cite{LEJ} 
in combination with the Argonne $V_{18}$ 
\cite{SCHI} as 
two-body component. In Ref.~[15] it was discussed the EOS of pure neutron 
matter 
and a preliminary prediction fos given for the symmetry energy based on 
the shift between 
the energy of pure neutron matter and symmetric nuclear matter. But one may 
argue that 3BF could modify the $\beta^2$ law fulfilled by the binding energy 
of asymmetric nuclear matter specially at high density and higher order terms
in the $\beta$ expansion could significantly limit the role played by
the symmetry energy in describing the isospin effects. All that required to  
to extend the calculation of the EOS to the full range $0 < \beta < 1$ in a 
baryonic density range high enough.    
In this paper we present the results of such a calculation. After describing the
model of TBF (Sec.~2) we will focus on the 3BF effects on binding energy, 
symmetry energy and single-particle properties
of asymmetric nuclear matter (Sec.~3). Then a few direct applications in the 
neutron star cooling and in the isospin properties of HIC will be shortly 
discussed along with a comparison with other approaches (Sec.~4).
    
{\bf 2.~Isospin dependent 3BF}

The microscopic 3BF adopted in the present calculation is based on the 
meson-exchange current approach.It is described in full detail in 
Ref.~[11]. The new meson parameters calculated to meet the 
self-consistent requirement with the adopted $AV_{18}$ 2BF are 
reported in Ref.~[13]. 

It has been shown \cite{MATH,LEJ} that the main contributions to 3BF 
rise from the two-meson exchange part of the NN interaction medium modified 
by the intermediate virtual excitation of nucleon resonances (isobar 
$\Delta (1232)$ and Roper $N(1440)$) (Fig.~1a) and from the two-meson exchange 
part with nucleon-antinucleon virtual excitation (Fig.~1d). This latter contains the
relativistic effects associated to the dressed spinors of the Dirac-Brueckner
approach \cite{MACH}. The terms associated to the non-linear $\pi$-nucleon coupling 
required by the chiral symmetry \cite{GLOC} play a minor role specially above 
the saturation density, where heavy mesons ($\sigma$ and $\omega$) are 
dominating over the $2\pi$-3BF (Fig.~1b, diagrams without heavy mesons). 
The contribution due to meson-meson coupling (Fig.~1c) is also negligible
\cite{MATH} and was not considered in the present calculation.
In the context of non relativistic approaches phenomenological 3BF have 
also been introduced\cite{PIEP} in the calculations of nuclear matter. 
They are modelled on the saturation properties of nuclear matter and/or 
binding energies of light nuclei. A good agreement between microscopic and
phenomenological 3BF is found only when this latter is treated within the 
Brueckner approach \cite{BBB}. 

The general Brueckner formalism for asymmetric nuclear matter (ANM) with 2BF's
is described in Ref.~[8]. The rigorous procedure to include 3BF would require
to solve the Bethe-Faddeev equation with a 3BF much the same as already done
to calculate the three-body clusters in the BBG hole-line expansion 
\cite{BALDO}.
But at this moment this appears a formidable task and we are content
to follow a simplified procedure based on converting the 3BF into an effective 
two-body force via a suitable integration over the third-nucleon degrees of 
freedom. The integral is weighted over the correlation function of the third
nucleon with the two others \cite{MATH,LEJ}. The effective 2BF for asymmetric
nuclear matter is defined as
\be
\begin{array}{lll}
 <\vec r_1 \vec r_2| V_3^{\tau_1\tau_2} | 
\vec r_1^{\ \prime} \vec r_2^{\ \prime} > 
&=&
\displaystyle
 \frac{1}{4} \sum_{\tau_3}\sum_{\sigma_3}\sum_n \int {\rm d}
{\vec r_3} {\rm d} 
{\vec r_3^{\ \prime}}\phi^*_n(\tau_3\vec r_3^{\ \prime})
(1-\eta_{\tau_1,\tau_3}(r_{13}' ))
(1-\eta_{\tau_2,\tau_3}(r_{23}')) \\[6mm]
&\times&
\displaystyle
W_3(\vec r_1^{\ \prime}\vec r_2^{\ \prime}
\vec r_3^{\prime}|\vec r_1 \vec r_2 \vec r_3)
\phi_n(\tau_3r_3)
(1-\eta_{\tau_1,\tau_3}(r_{13}))(1-\eta_{\tau_2,\tau_3}(r_{23}))
\end{array}\ee
The function $\eta_{\tau_1\tau_2}(r)$ is the average over spin 
and momenta in the Fermi sea
of the defect function, of which only the most important partial wave
components have been included, i.e. the $^1S_0$ and $^3S_1$ partial waves.

The transformation of the 3BF to an effective 2BF entails a 
selfconsistent coupling between  3BF and Brueckner procedure of solving
the Brueckner-Bethe-Goldstone equations. One first calculates the correlation function with only
the 2BF and then builds up the effective 3BF which in turn is added to the 2BF, 
and again calculates the correlation function and so on up to the convergence 
is reached. As previously mentioned the bare 2BF adopted in the calculations 
was the charge dependent $AV_{18}$. The partial wave expansion of 
full interaction has been truncated at $l_{max}=6$.   

{\bf 3.~ Numerical results}

The EOS of ANM has been calculated spanning the
whole asymmetry range with a step-size $\Delta\beta=.2$ and a density 
domain up to 0.45fm$^{-3}$. The case of symmetric nuclear 
matter ($\beta=0$) is discussed in Ref.~[15] (the saturation 
properties are reported in Table 1). 
The results for ANM are displayed in Fig.~2 for both 
cases with (left panel) and without (right panel) the 3BF. In the figure
energy shift for asymmetric-to-symmetric nuclear matter is reported versus
$\beta^2$. The individual runs (symbols in the Figure) are 
depicted along with their linear fits (solid lines) performed with 
only the first three values of asymmetry parameter $\beta$. From the comparison
between the two sets of calculations one may notice that, despite its strongly 
density dependent repulsive effect, the 3BF does not violate the $\beta^2$ law 
fulfilled already with only 2BF. 
The rather good agreement between the symbols and the corresponding lines 
(maximum deviation is $6\%$ ) indicates the high quality of the $\beta^2$ law
up to the largest densities. This is a quite astonishing result because
the 3BF introduces a strong density and isospin dependence, making the 
nucleon-nucleon ($NN$)
interaction quite different from the pure 2BF. On the other hand, 
it is a quite desirable result indeed for two reasons. 
First, this (in agreement with previous studies\cite{BOMB,ASI,KUO,WEB} 
using alone charge-independent 2BF) provides a strong 
support to the applications based on the EOS of $AMN$ 
extracted from the empirical $\beta^2$--law.
Second, it imposes also strict theoretical  
constrains on the phenomenological nuclear forces when extended to 
$AMN$. We have in mind the Skyrme forces which have been fit around the 
saturation point 
of symmetric nuclear matter and include an effective 
density-dependent term to simulate the effects of 3BF. 

In Table 1 the saturation properties of ANM are shown with and without 3BF.
One may just notice that the compression modulus results smaller 
with 3BF  despite the strong enhancement of the curvature of the EOS. 
This is due to the fact that it is also proportional to 
the square of the saturation density, which turns out to be very much reduced.  

The symmetry energy is defined as
\be
E_{sym}(\rho) \quad = \quad {1\over 2} \left[ {\partial^2 B(\rho,\beta)
\over \partial \beta ^2}\right]_{\beta=0}.
\ee
Due to the simple $\beta^2$-law the symmetry energy can be 
equivalently calculated as the difference between the energy per nucleon 
of pure neutron matter and symmetric nuclear matter, i.e., 
\be
E_{sym}(\rho) \,=\, E_A(\rho,1) - E_A(\rho,0). 
\ee
Fig.~3  shows the effect of the 3BF on the symmetry energy in 
the density domain considered in this study.
At the saturation density the two values do not significantly 
differ: 30.71 MeV (3BF included) and 29.28 MeV (no 3BF). Both are in good agreement 
with the empirical value $30. \pm 4$ MeV extracted from the nuclear mass
table~\cite{HAUST}. Above $\rho_0$ the 3BF gives a strong enhancement of 
symmetry energy since it is strongly repulsive  at high density. 
Both curves, with and without the 3BF, have been parametrized by simple 
power laws as follows,
\begin{itemize}
\item BHF with pure $AV_{18}$ 2BF\\
\be
E_{sym} = 30.7u^{0.58} 
\ee
\item BHF using $AV_{18}$ plus the 3BF\\
\be\begin{array}{llll}
E_{sym} & = & 30.71 u^{0.6}         & u \le 1 \\[3mm]
        & = & 30.71 + 18.42(u-1) + 9(u-1)^2  & u > 1 
\end{array}\ee
\end{itemize}
where $u=\rho/\rho_0$, and $\rho_0=0.17~$fm$^{-3}$ the empirical 
saturation density. The above simple  relations, plotted in Fig.~3 
(right panel), may be useful in HIC simulations. 
\par
In Fig.~3 (left panel) $E_{sym}$ vs density is compared with other 
approaches. Despite the overall agreement with the predictions of both 
relativistic mean-field (RMF) theory ~\cite{DITO1} and 
Dirac-Brueckner-Hartree-Fock 
approach (DBHF)~\cite{KUO}, the density dependence is found to be 
rather different. Both RMF and DBHF theories predict an almost linear variation 
of $E_{sym}$ vs. density. Instead, the present BHF calculation with the 3BF 
gives a slower varying 
$E_{sym}$ vs. $\rho$ at relatively low densities, say from 
$\rho\simeq 0.03$fm$^{-3}$ to $\rho\simeq 0.25$fm$^{-3}$. 
Such a slow
variation vs. density is also found in a density region up to
$\rho\simeq 0.2$fm$^{-3}$ in Ref.~\cite{KAIS} from the three-loop
approximation of chiral perturbation theory.
In this density region, the shape of $E_{sym}$ plays an important
role as to the study of the isospin effects in HIC at intermediate energy as
discussed later.
On the contrary, in the relatively high density domain, 
i.e., $\rho\ge 0.3$fm$^{-3}$, the present calculation 
predicts a steeper density dependence of $E_{sym}$ than DBHF and 
RMF.

Due to the isospin effect  the proton and neutron single-particle  
potentials  are different. As discussed in Ref.~\cite{ASI} where only the charge-independent  
$AV_{14}$ 2BF was used, the attractive $T=0$ $SD$ channel
contribution in the two-body NN interaction mainly drives the 
isospin dependence of the proton and neutron mean fields. 
As a result, the proton mean field becomes more attractive while 
the neutron one more repulsive as increasing asymmetry. 
In the BHF  approximation both mean fields  vary  linearly with the 
asymmetry parameter $\beta$,  which is in keeping with
the $\beta^2$ dependence of the energy per particle. 
A potential linearly dependent on $\beta$  was introduced phenomenologically 
long ago~\cite{LANE} and is referred to as Lane potential.
Due to the importance of the single-particle properties such as mean field 
and effective mass in HIC physics~\cite{BERT,CASS}, 
it is of some interest to explore 
the effect of the 3BF on such quantities. The proton and 
neutron mean fields are 
shown as a function of momentum $k$ at different asymmetries 
$\beta=0, 0.2, 0.4, 0.6, 0.8$ in Fig.~4 for the saturation density 
$\rho=0.17$fm$^{-3}$ and in Fig.~5 for high density $\rho=0.34$fm$^{-3}$. 
In both figures the left panels display the results using $AV_{18}$ plus 
the 3BF and the right panels those using pure $AV_{18}$. 
As expected, the 3BF adds a repulsive contribution to 
both proton and neutron mean fields at all asymmetries. 
At relatively low density (Fig.~4), the proton mean field 
with 3BF becomes more attractive, which is in agreement with the 2BF 
prediction.
However the validity of the linear Lane assumption is  
broken by 3BF as more clearly seen in the 
left panel of Fig.~6, where the isospin variation of the 
proton and neutron mean fields are plotted at momentum $k=0$. 
At relatively high density (Figs.~5 and 6), the 3BF force effect 
becomes much more pronounced and in fact it brings a strong deviation from 
the linear Lane assumption. The neutron mean field rises up 
more rapidly as compared to the results with pure $AV_{18}$. The 
same happens to the proton potential which at a certain asymmetry becomes 
even more repulsive. 
. This remarkable result can be 
explained by the competition between the isospin dependence 
of the 3BF and the contribution from the attractive 
$T=0$ $SD$ channel. As increasing isospin asymmetry
the 3BF repulsion starts to compete with the $T=0$ $SD$ channel 2BF attraction, 
and becomes the dominant one at density high enough. 

The effective mass, which is related to the non-locality (momentum 
dependence) of the neutron and proton mean fields, is defined as 
\be
{ m^*_{\tau}(k) \over m} \, = \, {k\over m}
\left( {dE^{\tau}(k)\over{dk}} \right)^{-1}.
\ee
The momentum dependence of $m^*$ is featured by a wide bump 
inside the Fermi sphere due to the high probability amplitude 
for particle-hole excitations near the Fermi surface\cite{MAHA0}.
The isospin dependence of the 
proton and neutron effective masses at their respective Fermi 
momentum $k_F^p$ and $k_F^n$ are given in the right panel of 
Fig.~(6). The 3BF does not affect the linear 
scissor-shaped behavior observed in the previous calculations using a pure 
2BF~\cite{BOMB,ASI}. One should notice that the isospin effect on neutron and 
proton effective masses in the Brueckner approach goes the other way around   
than in the RMF approach \cite{DITO1}. This discrepancy is to be understood taking
also into account the different definitions of effective mass in the two 
approaches.
  
{\bf 4.~Summary and discussion}

In the present work the microscopic 3BF based on meson-exchange current 
approach has been extended and applied to isospin ANM 
in the framework of the Brueckner theory. The 3BF effects 
on the isospin dependence of both the nuclear EOS and single-particle 
properties have been investigated. 

The results 
confirm the validity of the $\beta^2$ law for  the energy per nucleon
in the entire range of isospin asymmetry and up to high density 
in spite of the strong isospin and density dependence of the 
3BF. As a consequence the isospin effects in ANM are just driven by the 
symmetry energy as in the case with only the 2BF.  
The vanishing of higher powers in the $\beta^2$ expansion also supports   the 
simple recipe often adopted to extract the symmetry energy 
 from the two limiting cases of symmetric nuclear matter and pure neutron 
matter . 
It also constraints theoretically the phenomenological interactions, 
such as the Skyrme forces which take into account the effect of 3BF by a 
density dependent term. 

As expected, the 3BF improves the saturation 
properties of symmetric nuclear matter by shifting the equilibrium 
density close to the empirical value. At relatively low density, the 3BF effect 
on the nuclear symmetry energy is quite small. On the contrary, 
at high density, it brings a strong enhancement and consequently the symmetry 
energy rises with density more steeply  than the corresponding 2BF 
prediction. 
The non linear increase of symmetry energy at high density has also been observed in a recent 
relativistic Hartree-Fock calculation~\cite{DITO1} as the `Fock' exchange 
effect of the non linear scalar self-interactions. 
But in the same Ref.~\cite{DITO1}, it is also shown that at sub-nuclear densities, 
the Fock contribution could result in a softening of the symmetry potential 
term.

The density dependence of the symmetry energy 
has been parametrized for the sake of application in HIC with very neutron-rich
ions.  It has been shown already that isospin 
fractionation~\cite{BAO2,DITO2} in  multifragmentation events is very 
sensitive to the density dependence of $E_{sym}$ in the low density region.
In particular, in Ref.~\cite{BAO3} is shown that using an isospin stiff 
nuclear EOS with symmetry energy curvature
equal to  -69 MeV (very close to the present value of -66 MeV)
leads to a remarkable value for the liquid-to-gas isospin asymmetry ratio
which is consistent with the experimental prediction \cite{XU}. 
An isospin scaling has been proposed in multifragmentation events of HIC,
which turns out to be also very sensitive to the density dependence of 
$E_{sym}$. Using the expanding evaporating source model and adopting for $E_{sym}$ the 
simple parametrization $ C \cdot (\rho/\rho_0)^{\gamma}$ the 
fragment data can be fairly well reproduced by taking $\gamma=0.6$ \cite{LYNCH}
, which coincides with the fit of our microscopic value (cfr. Eq.~(5)).

The preequilibrium particle 
emission~\cite{BAO,BAO1} and collective flows~\cite{BAO2,DITO3} can instead 
probe the symmetry energy in the range of  high density, say up to two times the 
saturation density, but only with an isospin stiff EOS which is in agreement 
with our prediction.

 In the high density range the symmetry energy is also relevant to the study of 
the neutron-star properties such as the cooling mechanism. In fact, a steep 
increase of $E_{sym}$ with density favours the direct 
URCA processes~\cite{LATT}. In Fig.~7 the proton fraction is reported for
 $\beta$-equilibrium  nuclear matter in different approximations. The 
threshold of direct URCA processes is only crossed by the results with 3BF at
a reasonable value of the neutron-star density.   
 
As to the single-particle properties, the effect of the 
3BF is to add an extra repulsion to both proton and neutron 
mean fields so that the linear Lane assumption breaks down slightly around the 
saturation 
density but quite strongly at high density. However, the scissor-shaped 
behavior  of the proton 
and neutron effective masses vs. $\beta$ remains unchanged being the momentum 
dependence of the mean fields rather insensitive to 3BF. 
In a calculation not reported here it was found that the 3BF affects only 
slightly the {\it rearrangement} contribution to the nuclear mean field
and can not improve the fulfillment of Hughenoltz-Van Hove theorem 
(see Ref.~[8]). This requires to go beyond BHF approximation in the expansion 
of mass operator that is a work still in progress. 


\vskip 1cm
{\bf Acknowledgments:} \\

  One of us (W.~Z.) would like to thank INFN-LNS and
  the Physics Department of the Catania University (Italy)
  for their hospitality during the preparation of the present work.

  This work has been supported in part by the Chinese Academy of Science,
  within the {\it one Hundred Person Project}, the Major State Basic
  Research Development Program of China under No.~G2000077400.


\newpage
{\footnotesize{Tab.~1 - Equilibrium density of asymmetric 
nuclear matter, incompressibility and energy per nucleon at 
equilibrium density corresponding to four different values 
of asymmetry $\beta$. }
\\
\begin{center}
\begin{tabular}{ l l l l l l l} \hline\hline
& \multicolumn{3}{c}{$AV_{18} + 3BF$} & 
\multicolumn{3}{c}{$AV_{18}$}  \\ \hline\hline
$~~\beta$ ~~~~~~~~ & $\rho_{eq}~~~~~$ & $K(\rho_{eq})~~$ 
& $E_A(\rho_{eq})$ ~~~~~~~~~& 
 $\rho_{eq}$~~~~~ & $K(\rho_{eq})$~~& $E_A(\rho_{eq})$ 
\\ \hline
 ~~0.0 &0.198 &207.84 &$-15.05$ & 0.265 &232.35 &$-18.25$ \\
 ~~0.2 &0.193 &195.34 &$-13.82$ & 0.259 &228.54 &$-16.74$ \\
 ~~0.4 &0.165 &142.33 &$-9.93$  & 0.226 &177.49 &$-12.38$ \\
 ~~0.6 &0.120 & 85.28 &$-4.44$  & 0.172 & 96.1  & $-5.89$ \\ 
\hline \hline
\end{tabular}
\end{center}
\begin{figure}
\caption{Diagrams of the microscopic 3BF adopted in the
 present calculation (see Ref.~[11]). Diagram (c) was not included.}
\end{figure}
\begin{figure}
\caption{Energy per nucleon of asymmetric nuclear 
matter in the range $0\le \beta^2 \le 1$ at 
four densities as compared to 
the parabolic fits ( straight lines ) obtained from the 
first three values of $\beta$ (0.0, 0.2, 0.4). \ 
Left panel : BHF predictions using $AV_{18}$ plus the 3BF. 
Right panel: BHF results with only pure $AV_{18}$ 2BF.} 
\end{figure}
\begin{figure}
\caption{ Symmetry energy vs density. The left panel shows the comparison 
among different approaches: the BHF predictions with 3BF (upper solid curve)
and without 3BF (lower solid curve) are obtained from the slopes of Fig.~2 
(while symbols are the values approximated by Eq.~(3)); 
the long-dash curve corresponds to the result of DBHF approach from 
Ref.~[20]. The short dash is that of BHF calculation using 
the phenomenological Urbana 3BF in Ref.~[19], where $E_{sym}$ is 
obtained from Eq.~(3). The dotted curve is the prediction of relativistic 
Hartree-Fock approach taken from Ref.~[23].
Right panel shows simple parametrizations of the present results 
with (solid line) and without (dashed line) 3BF.} 
\end{figure}
\begin{figure}
\caption{ Proton and neutron mean fields 
in asymmetric nuclear matter 
at $\rho=0.17 fm^{-3}$ for five different asymmetries.
Left part shows the proton mean-field (upper panel) and the neutron 
one (lower panel), respectively, vs. momentum using $AV_{18}$ plus 
the 3BF. Right part shows the corresponding results without the 3BF.}
\end{figure}
\begin{figure}
\caption{ The same as in Fig.~4 for $\rho=0.34$fm$^{-3}$. }
\end{figure}
\begin{figure}
\caption{ Left panel: Isospin variation of proton and neutron 
mean fields at a fixed momentum $k=0$ for two densities 
$\rho=0.17$ and 0.34fm$^{-3}$. 
Right panel: isospin dependence of proton and neutron 
effective masses calculated at their respective Fermi momenta. 
The results in both panels are calculated using $AV_{18}$ plus the 3BF. }
\end{figure}
\begin{figure}
\caption{Proton fraction in $\beta$-equilibrium nuclear matter calculated
in the BHF approximation with 2BF ($AV_{18}$) and 2BF plus 3BF ($AV_{18}+3BF$)
in comparison with a variational calculation using Urbana 2BF and 3BF 
($UV_{14}+UVII$). The dashed orizontal line is the threshold for direct URCA 
processes estimated in Ref.[34]}  
\end{figure}
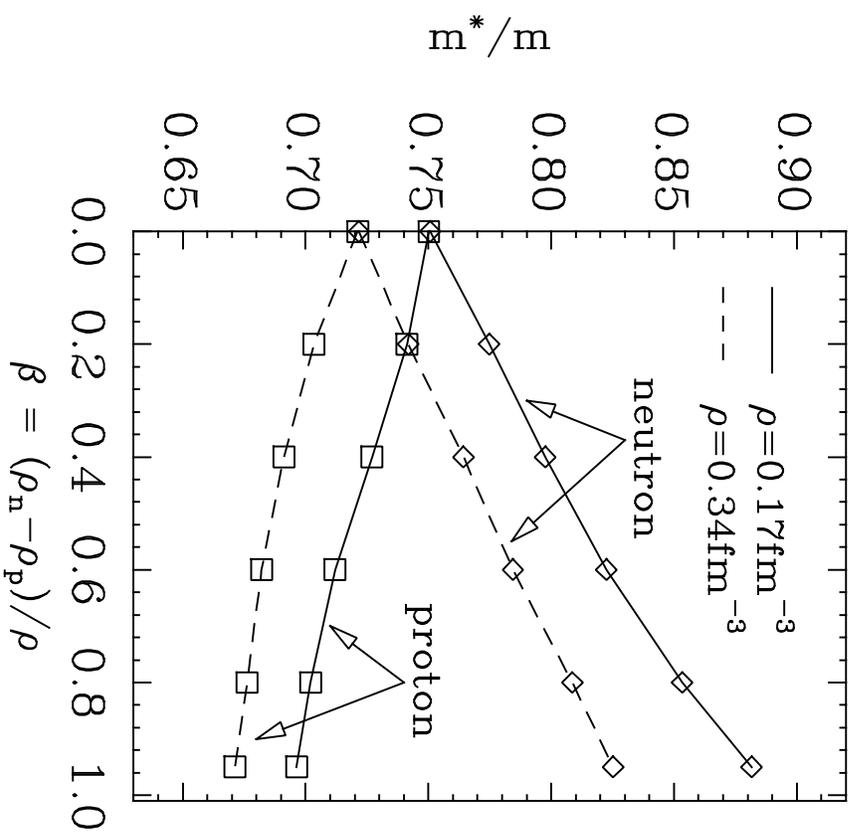
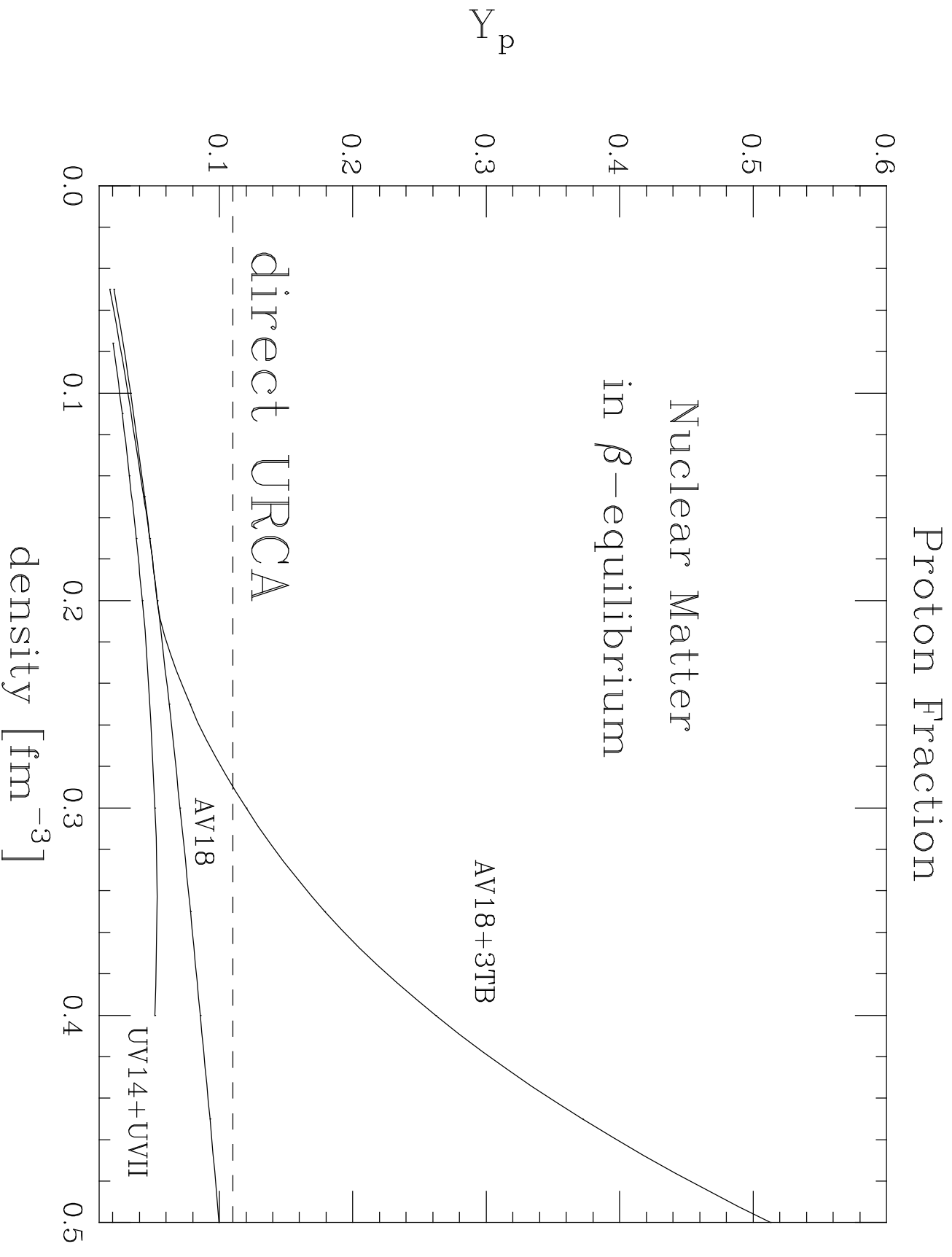
\end{document}